\documentclass[12pt]{article}

\usepackage{amsmath, amssymb, amsfonts}
\usepackage{graphicx}
\usepackage{tikz}
\usetikzlibrary{positioning, arrows.meta, shapes.multipart}
\usepackage{pgfplots}
\usepackage{booktabs}
\usepackage{geometry}
\usepackage{hyperref}
\usepackage{cite}
\usepackage{pgfplots}
\pgfplotsset{compat=1.17}
\pgfplotsset{compat=1.18}

\title{\textbf{Inverse Hamiltonian Reconstruction from Gravitational Energy Density in Curved Spacetime}}
\author{Davood Momeni \\
\small Department of Physics, Northeast Community College, Norfolk, NE 68701, USA\\
\small Centre for Space Research, North-West University, Potchefstroom 2520, South Africa}
\date{}

\begin{document}

\maketitle
\begin{abstract}
We present a general framework for reconstructing effective Hamiltonians from known gravitational energy density profiles in curved spacetime. Starting from local thermal equilibrium and Liouville dynamics, we establish an inverse procedure that relates the macroscopic energy density \( \rho(x) \) to a distribution function \( f(x,p) \sim e^{-\beta H(x,p)} \), and recovers the underlying Hamiltonian \( H(x,p) \) via functional inversion. This approach synthesizes tools from relativistic kinetic theory, statistical mechanics, and covariant gravitational thermodynamics, offering a systematic way to extract microscopic dynamics from coarse-grained energy observables. Applications include FLRW cosmology, Loop Quantum Gravity corrections, AdS/CFT holography, and the SYK model. Our results provide a novel route for probing emergent spacetime dynamics through observable densities, bridging geometry, entropy, and Hamiltonian flow in curved backgrounds.
\end{abstract}

\vspace{1em}


\section{Introduction}
Hamiltonian dynamics lies at the heart of classical and quantum physics, dictating the evolution of physical systems through phase space. In gravitational theories, especially those involving curved spacetimes, the role of Hamiltonians becomes obscured due to coordinate dependence, gauge freedom, and the non-local nature of gravitational energy. Nevertheless, in thermodynamic and cosmological settings, one often has access to energy density functions $\rho(x)$ derived from either observations or theoretical models.

In curved spacetime, defining local Hamiltonians is intrinsically complicated by the absence of global time-like Killing vectors and the ambiguity in localizing gravitational energy. This makes direct formulation of dynamics in terms of $H(x,p)$ problematic. However, energy density profiles $\rho(x)$, often derived from observational cosmology or black hole thermodynamics, remain well-defined in specific frames. Inverting such energy data to reconstruct the underlying dynamical generator thus offers a compelling indirect route to accessing the microscopic physics of gravitational systems.

This leads to a natural but challenging inverse problem: \textit{Given a gravitational energy density function, can one reconstruct the effective Hamiltonian governing the underlying dynamics}? This question is relevant in early universe cosmology, emergent gravity scenarios, and holography. It also resonates with recent approaches in statistical mechanics and Liouville dynamics, where macroscopic densities are tied to microscopic dynamics via distribution functions.

The connection between macroscopic energy densities and microscopic dynamics is naturally encoded in the Liouville equation, which governs the conservation of probability in phase space. When combined with the assumption of local thermodynamic equilibrium, this framework allows one to express $\rho(x)$ as an integral over distribution functions involving the Hamiltonian. The reverse problem—extracting $H(x,p)$ from known density data—requires careful treatment of this relation and motivates the inversion method developed in this work.

We approach this reconstruction problem by embedding it in a framework that unifies kinetic theory, energy conservation in gravitational systems, and generalized equilibrium assumptions. In particular, we derive a formal inverse map from $\rho(x)$ to $H(x,p)$ under thermal assumptions $f \sim e^{-\beta H}$, and illustrate its implications in Friedmann-Lemaître-Robertson-Walker (FLRW) cosmology, Loop Quantum Gravity, AdS/CFT correspondence, and the SYK model.

In modified theories of gravity, such as those involving non-minimal couplings or torsion, the conservation of the energy-momentum tensor is generally violated. This implies that the standard forward Hamiltonian formulation may not directly apply. Nevertheless, if the matter content is approximately in equilibrium, Liouville-based statistical approaches remain valid, and an effective Hamiltonian can still be inferred from observationally determined energy densities, as we explore in this framework.

Our inversion framework resonates with themes in holography and emergent gravity, where boundary observables are often viewed as encodings of deeper bulk dynamics. In this context, energy density on a spatial hypersurface can be interpreted as a coarse-grained imprint of an underlying Hamiltonian structure. Reconstructing this structure from macroscopic data offers a novel route to explore the dynamical content of spacetime and quantum fields, independent of any assumed Lagrangian or metric ansatz.

To formulate this reconstruction precisely, we develop a general framework that synthesizes tools from relativistic kinetic theory, gravitational energy flow, and equilibrium statistical mechanics. We construct an inverse map from the spatial energy density $\rho(x)$ to the effective Hamiltonian $H(x,p)$ under the local thermal ansatz $f(x,p) \sim e^{-\beta H(x,p)}$. Applications of this method are explored in several physical contexts, including FLRW cosmology, Loop Quantum Gravity corrections, AdS/CFT dualities, and the Sachdev-Ye-Kitaev (SYK) model.

\section{Covariant Energy Conservation in Gravity}

In General Relativity (GR), the energy-momentum tensor $T^{\mu\nu}$ satisfies the local conservation law $\nabla_\mu T^{\mu\nu} = 0$, which is a direct consequence of diffeomorphism invariance of the Einstein-Hilbert action. This law ensures that energy and momentum are conserved in curved spacetime, albeit in a covariant (rather than coordinate-invariant) sense.

However, in generalized theories of gravity—especially those that extend the Levi-Civita connection to include torsion or non-metricity—this conservation law is generically modified. In particular, in metric-affine theories, the divergence of the energy-momentum tensor acquires a non-zero source term:
\begin{equation}
\nabla_\mu T^{\mu\nu} = \mathcal{J}^\nu,
\end{equation}
where $\mathcal{J}^\nu$ is a geometrically induced current reflecting the presence of additional gravitational degrees of freedom. This current may originate from torsional spin couplings, trace anomalies, or energy exchange with geometric structures like non-metricity vectors or scalar connections \cite{Hehl1995, BeltranJimenez2019}.

To derive a continuity-type equation for the energy density, we project the divergence along the four-velocity $u^\mu$ of a comoving observer. The contraction $u_\nu \nabla_\mu T^{\mu\nu}$ yields:
\begin{equation}
u_\nu \nabla_\mu T^{\mu\nu} = \frac{d\rho}{d\tau} + \rho \nabla_\mu u^\mu + \cdots,
\end{equation}
where $\rho = T_{\mu\nu} u^\mu u^\nu$ represents the local energy density as measured by the observer, and $\tau$ is the proper time along their worldline.

The second term, $\rho \nabla_\mu u^\mu$, describes the dilution or amplification of energy due to expansion or contraction of spacetime (analogous to fluid divergence). Additional terms, omitted here for brevity, include pressure gradients, anisotropic stresses, or energy fluxes orthogonal to $u^\mu$.

This formulation leads to a generalized gravitational continuity equation, which underpins many phenomena in cosmology and astrophysics. In particular, it bridges gravitational field equations with thermodynamic evolution equations, and plays a crucial role in deriving entropy balance laws in both classical and semiclassical contexts. As we show in the following sections, this framework allows for a geometric reinterpretation of energy flow and opens a pathway to reconstruct Hamiltonian structures from macroscopic observables.
\section{Liouville Equation and Distribution Function}

In relativistic kinetic theory, the state of a dilute system of particles is described by a one-particle distribution function $f(x^\mu, p^\nu)$ on phase space, where $x^\mu$ and $p^\nu$ denote spacetime coordinates and momenta, respectively. This function gives the expected number of particles in a differential volume of phase space and evolves according to the Liouville—or Vlasov—equation \cite{Degroot1980,Ehlers1971}:
\begin{equation}
\frac{df}{d\tau} = \frac{p^\mu}{m} \frac{\partial f}{\partial x^\mu} - \Gamma^\lambda_{\mu\nu} \frac{p^\mu p^\nu}{m} \frac{\partial f}{\partial p^\lambda} = C[f],
\end{equation}
where $m$ is the rest mass of the particles, $\Gamma^\lambda_{\mu\nu}$ are the Christoffel symbols representing gravitational effects, and $C[f]$ is the collision term. In the collisionless (Vlasov) regime, $C[f] = 0$, and the distribution function evolves strictly under geodesic flow, preserving phase-space density.

This equation expresses the fact that the number of particles is conserved along their trajectories, and it geometrically encodes how spacetime curvature affects the evolution of matter. In the Hamiltonian formulation of classical mechanics, this is an expression of Liouville's theorem: the phase-space volume element is conserved along the flow generated by the Hamiltonian vector field. Thus, even in a curved background, the evolution is incompressible in the symplectic sense.

Assuming equilibrium or near-equilibrium, the distribution function typically takes the Boltzmann-Gibbs form:
\begin{equation}
f(x^\mu, p^\nu) \sim e^{-\beta(x) H(x,p)},
\end{equation}
where $\beta(x) = 1/T(x)$ is the inverse temperature field, and $H(x,p)$ is the effective Hamiltonian describing the energy of a single particle in a local inertial frame. This exponential form arises from the principle of maximum entropy and is consistent with both microcanonical and canonical ensembles in the appropriate limit.

Importantly, $H(x,p)$ is not necessarily known a priori in gravitational systems, especially in generalized or emergent gravity scenarios. Instead, what is often known is the macroscopic profile of $f$ or the derived energy density $\rho(x)$. Therefore, the Liouville equation provides not only a predictive dynamical tool but also a diagnostic framework: by analyzing how $f$ behaves in spacetime, one may infer properties of the underlying Hamiltonian or effective spacetime geometry.

This viewpoint offers a powerful bridge between statistical physics and spacetime dynamics. It underlies much of modern cosmological perturbation theory, kinetic approaches to dark matter, and semi-classical gravity. In this work, we build upon this connection to solve the inverse problem: recovering $H(x,p)$ from $\rho(x)$ under suitable assumptions on $f$. This inversion is guided by the equilibrium form above and the thermodynamic structure implied by the Liouville flow.

\section{Inverse Problem: From $\rho$ to $H$}

In many physical and gravitational systems, the energy density $\rho(x)$ is a macroscopic observable, derived either from averaging microscopic matter distributions or via geometric contributions such as in the ADM or Komar formulations. However, in kinetic theory and Hamiltonian mechanics, the energy density is fundamentally expressed in terms of the microscopic Hamiltonian and a distribution function:
\begin{equation}
\rho(x) = \int H^2(x,p)\, f(x,p)\, d^3p.
\end{equation}
Here, \( H(x,p) \) represents the effective energy of a particle at position \( x \) with momentum \( p \), and \( f(x,p) \) encodes the statistical weight of such microstates. The integral over momentum space collapses the microstructure into a coarse-grained scalar quantity.

Our central goal is to invert this relation: to obtain the Hamiltonian \( H(x,p) \), or at least an approximation to it, from the known macroscopic profile \( \rho(x) \). This constitutes a non-trivial inverse problem, especially when the distribution function is unknown or only partially constrained.

We proceed by assuming thermal equilibrium, where the distribution follows the Boltzmann form:
\begin{equation}
f(x,p) \propto e^{-\beta(x) H(x,p)},
\end{equation}
with \(\beta(x) = 1/T(x)\) being the inverse local temperature field. This assumption, though idealized, is justified in many contexts where thermalization occurs, such as in cosmological eras following inflation or in quasi-static black hole configurations.

Inserting this form into the energy density expression and performing a saddle-point or variational analysis, one can derive an approximate inverse relation:
\begin{equation}
H(x) \sim -T(x) \log \rho(x) + \text{const}.
\end{equation}
This equation captures the essence of an effective Hamiltonian encoded in the entropy content of the density. It bears resemblance to Legendre duality in thermodynamics and to entropy regularization techniques used in variational inference and optimal transport.

The above formula is particularly powerful in systems where \( \rho(x) \) exhibits strong geometric dependence, such as in warped spacetimes or in theories with holographic duals. It provides a direct route to extract dynamical information (via \( H \)) from observational or coarse-grained data (via \( \rho \)).
\begin{figure}[htbp]
\centering
\begin{tikzpicture}[
    node distance=1.5cm and 0cm,
    block/.style={
        rectangle,
        draw=black,
        rounded corners,
        text width=4cm,
        align=center,
        minimum height=1.2cm,
        font=\footnotesize,
        fill=#1
    },
    arrow/.style={thick,->,>=Stealth}
]

\node[block=cyan!20] (rho) {Input\\$\rho(x)$ known};
\node[block=yellow!20, below=of rho] (assume) {Assume\\$f(x,p) \propto e^{-\beta H}$};
\node[block=orange!20, below=of assume] (relation) {Relation\\$\rho = \int H^2 f\, d^3p$};
\node[block=green!25, below=of relation] (invert) {Approximate Inversion\\$H \sim -T \log \rho$};
\node[block=red!20, below=of invert] (output) {Output\\$H(x)$ recovered};

\draw[arrow] (rho) -- (assume);
\draw[arrow] (assume) -- (relation);
\draw[arrow] (relation) -- (invert);
\draw[arrow] (invert) -- (output);

\end{tikzpicture}
\caption{Inverse map from energy density to effective Hamiltonian using equilibrium assumption.}
\label{fig:inverseH}
\end{figure}
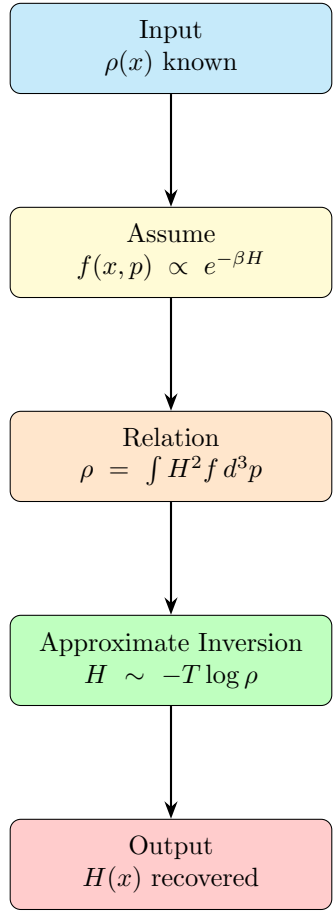

\section{Conditions for Equilibrium Ansatz in Curved Spacetime}

The equilibrium ansatz for the distribution function,
\begin{equation}
f(x, p) \propto \exp\left(-\beta H(x, p)\right),
\end{equation}
is valid under certain geometric and physical assumptions. In curved spacetime, this form emerges naturally when the system is in local thermodynamic equilibrium and the background admits a well-defined timelike Killing vector field \(\xi^\mu\). The condition
\[
\mathcal{L}_\xi g_{\mu\nu} = 0
\]
guarantees stationarity, ensuring that physical quantities are time-independent in the frame defined by \(\xi^\mu\). In such cases, the inverse temperature \(\beta\) can be associated with a redshifted Tolman factor:
\[
\beta(x) = \beta_0 \sqrt{-\xi^\mu \xi_\mu},
\]
as originally proposed in Tolman's law of relativistic thermodynamics.

Even in the absence of a strict Killing symmetry, the ansatz may remain approximately valid in a near-equilibrium setting if the relaxation time is much shorter than the spacetime curvature scale, as described in Israel-Stewart theories and Carter's variational formulation of relativistic fluids.

We emphasize that this ansatz corresponds to the zeroth-order solution of the Boltzmann equation under the relaxation-time approximation and is compatible with local thermodynamic equilibrium. It has been used extensively in studies of kinetic theory in general relativistic settings (e.g., \cite{CercignaniKremer, Ehlers1971}).
\section{Covariant Structure and Frame Dependence of the Reconstruction}

The inversion formula derived in this work, where the effective Hamiltonian \(H(x,p)\) is reconstructed from the spatial energy density \(\rho(x)\), inherently involves a foliation of spacetime into spatial hypersurfaces and a choice of local thermal frame. In general relativity, this is equivalent to selecting a timelike vector field \(u^\mu(x)\) corresponding to the local rest frame of the matter or observer.

To address coordinate dependence, we note that both \(\rho(x)\) and \(H(x,p)\) can be defined on a chosen hypersurface \(\Sigma_t\) with induced metric \(h_{ij}\), using the Arnowitt–Deser–Misner (ADM) formalism. In this approach, the Hamiltonian becomes a scalar under spatial diffeomorphisms, and its time evolution is governed by the lapse function \(N(x)\) and shift vector \(N^i(x)\), allowing us to write
\[
H(x,p) = N(x) \mathcal{H} + N^i(x) \mathcal{H}_i,
\]
where \( \mathcal{H} \) and \( \mathcal{H}_i \) are the Hamiltonian and momentum constraints, respectively.

Alternatively, in a locally inertial frame at a point \(x\), the vielbein \(e^\mu_a(x)\) allows the decomposition
\[
H(x,p) = H\left(e^\mu_a(x) p_\mu\right),
\]
and this formulation respects local Lorentz invariance by construction. The use of vielbeins thus provides a covariant representation of the reconstruction that reduces to the flat-space result in the Minkowski limit.

We also clarify that the inverse temperature \(\beta\) is not a global scalar in general spacetime, but transforms according to Tolman's law:
\[
\beta(x) = \beta_0 \sqrt{-g_{00}(x)} \quad \text{(in static coordinates)},
\]
or more generally using the norm of a timelike Killing vector when available.

Hence, while the inversion is not strictly coordinate-invariant, its geometrical structure can be cast covariantly via the ADM decomposition or the vielbein formalism, making it suitable for applications in curved spacetimes with a well-defined foliation.

\section{Example: FLRW Cosmology}

To illustrate the inverse reconstruction of the Hamiltonian from a known energy density profile, let us consider the Friedmann-Lemaître-Robertson-Walker (FLRW) cosmological model. The FLRW spacetime describes a homogeneous and isotropic universe, supported by strong observational evidence from the cosmic microwave background and large-scale structure surveys 
\cite{EllisMaartensMacCallum}-\cite{Planck2018}.

The metric for a spatially flat FLRW universe is:
\begin{equation}
ds^2 = -dt^2 + a^2(t) dx_i dx^i,
\end{equation}
where \( a(t) \) is the cosmic scale factor and \( t \) is the cosmic time. This geometry admits a natural foliation into spacelike hypersurfaces with vanishing spatial curvature, simplifying many calculations while still capturing essential features of cosmic expansion.

In such a spacetime, consider a gas of relativistic particles described by a distribution function \( f(p, t) \) that depends on the physical momentum \( p = |\vec{p}| \) and cosmic time \( t \). The effective Hamiltonian for a single particle is given by the usual special-relativistic energy:
\begin{equation}
H(p, t) = \sqrt{\frac{p^2}{a^2(t)} + m^2},
\end{equation}
where \( m \) is the rest mass of the particle, and the factor \( a^{-2}(t) \) arises from converting comoving momentum to physical momentum.

Assuming local thermal equilibrium at each time slice, the one-particle distribution function takes the Boltzmann form:
\begin{equation}
f(p, t) = A(t) \exp\left(-\beta(t) H(p, t)\right),
\end{equation}
where \( \beta(t) = 1/T(t) \) is the inverse temperature, and \( A(t) \) is a time-dependent normalization factor. This functional form corresponds to a comoving observer’s frame and is a valid approximation in the early universe where high interaction rates maintain local equilibrium.

The macroscopic energy density \( \rho(t) \) is then obtained by integrating over all particle states:
\begin{equation}
\rho(t) = \int H^2(p,t)\, f(p,t)\, d^3p = 4\pi \int_0^\infty H^2(p,t)\, f(p,t)\, p^2 dp.
\end{equation}
This expression captures how microscopic energy levels, weighted by their thermal occupation probability, collectively determine the average energy density of the fluid.

Our goal is to invert this integral relation to extract an approximate form of the Hamiltonian as a function of macroscopic observables. Although the full inversion is analytically intractable due to the nonlinear structure of \( H(p, t) \), one can approximate the result using asymptotic expansions or saddle-point approximations. For instance, in the ultrarelativistic regime where \( p \gg m a(t) \), the Hamiltonian simplifies to \( H(p, t) \approx p/a(t) \), and the integral can be evaluated in terms of gamma functions.

Assuming this limit and treating \( A(t) \) as slowly varying, we find that the energy density scales as:
\begin{equation}
\rho(t) \propto T^4(t),
\end{equation}
which is consistent with the Stefan–Boltzmann law for a radiation-dominated universe. Taking the logarithm and solving for the Hamiltonian leads to the inverse relation:
\begin{equation}
H(t) \sim -T(t) \log \rho(t) + \text{const}.
\end{equation}
This expression shows how one can reconstruct the effective Hamiltonian—or more precisely, its thermal scale—by measuring the time evolution of the energy density.

Physically, this relation encapsulates the idea that the expansion of the universe redshifts momenta and dilutes energy densities, and that the underlying particle dynamics must adjust accordingly to maintain thermal equilibrium. Inverting \( \rho(t) \) to obtain \( H(t) \) is thus equivalent to extracting the dynamical content of the cosmological fluid from its thermodynamic history.

This reconstruction method can also accommodate cases where the energy density evolves nontrivially due to entropy production, phase transitions, or modified gravity effects. By tracking how \( \rho(t) \) departs from standard scaling laws, one can probe deviations in the underlying Hamiltonian structure—potentially signaling the presence of new physics in the early universe.

In the following sections, we will generalize this procedure to incorporate quantum gravity corrections and holographic dualities, thereby demonstrating the universality and adaptability of the inverse Hamiltonian framework across gravitational contexts.

\begin{table}[h!]
\centering
\caption{Reconstruction of $H(t) \sim -T(t) \log \rho(t)$ in a radiation-dominated universe}
\begin{tabular}{cccc}
\toprule
$T(t)$ & $\rho(t)$ & $\log \rho(t)$ & $H(t)$ \\
\midrule
1.0   & 0.657               & $-0.420$   & 0.420 \\
0.5   & 0.041               & $-3.193$   & 1.597 \\
0.3   & 0.0054              & $-5.220$   & 1.566 \\
0.1   & $6.58 \times 10^{-5}$ & $-9.630$   & 0.963 \\
0.01  & $6.58 \times 10^{-9}$ & $-18.800$  & 0.188 \\
\bottomrule
\end{tabular}
\label{tab:Hfromrho}
\end{table}

\section{Extensions: Quantum Gravity and Holography}

While the classical Hamiltonian reconstruction from macroscopic energy densities is already insightful in relativistic and cosmological contexts, it becomes even more profound when extended to regimes where quantum gravitational effects become significant. In such scenarios, classical notions of spacetime and energy density often break down or become ambiguous due to quantum fluctuations, discreteness of geometry, or strong coupling effects. Nonetheless, effective energy profiles—whether in the form of expectation values of stress tensors, horizon entropies, or radiation fluxes—often remain accessible and can be used to infer aspects of the underlying microscopic dynamics. This opens a pathway to probe quantum corrections to classical Hamiltonians using observational or semi-classical data, thereby enriching our understanding of early universe physics, black hole interiors, and Planck-scale phenomena.

A particularly fruitful approach to this problem arises from holographic dualities, especially the AdS/CFT correspondence, where energy densities in the bulk spacetime correspond to field-theoretic operators on the boundary. In this framework, the gravitational energy profile encodes information about the dual Hamiltonian of the conformal field theory, allowing a reconstruction of quantum dynamics from geometric data. Similarly, in Loop Quantum Gravity, the discrete nature of space modifies the Hamiltonian constraint at the quantum level, leading to effective Hamiltonians with non-polynomial structure—yet still tied to energy densities via continuity-like relations. The Sachdev-Ye-Kitaev (SYK) model offers another perspective, where spectral densities and entropy functionals play roles analogous to gravitational observables. Together, these frameworks demonstrate that the inverse problem of Hamiltonian reconstruction is not merely a classical exercise, but a unifying concept that bridges semi-classical geometry, quantum dynamics, and statistical physics.

\subsection{Loop Quantum Cosmology}

Loop Quantum Cosmology (LQC) is a symmetry-reduced application of Loop Quantum Gravity (LQG) to homogeneous and isotropic spacetimes, particularly those described by the FLRW metric \cite{Ashtekar2006}-\cite{Corichi2011}. It introduces discrete quantum geometric effects into the early universe, most notably replacing the classical Big Bang singularity with a quantum bounce. The key modification comes from polymer quantization, where the gravitational phase space variables are no longer represented by standard position and momentum operators but by holonomies and fluxes.

In this formalism, the classical Hamiltonian constraint is replaced by an effective Hamiltonian that captures leading-order quantum corrections. For a flat FLRW model, the effective Hamiltonian for the geometry takes the form:
\begin{equation}
H \sim \frac{\sin(\lambda P)}{\lambda},
\end{equation}
where \( P \) is canonically conjugate to the scale factor \( a \), and \( \lambda \) is a parameter encoding the discreteness of quantum geometry—often associated with the minimum nonzero eigenvalue of the area operator in LQG. Unlike the classical relation \( H \sim P \), the sine function regularizes the Hamiltonian at high curvatures, preventing divergences and enabling a bounce scenario.

The energy density of the scalar field (or matter content) in this effective model is typically written as:
\begin{equation}
\rho(t) \sim \frac{H^2}{a^3(t)},
\end{equation}
indicating that the quantum gravitational energy is still linked to the effective Hamiltonian, though with nontrivial functional dependence due to the holonomy corrections. This expression is derived from semiclassical analysis and has been validated by comparing with full quantum evolution in LQC.

Given a known energy density profile \( \rho(t) \) (e.g., inferred from cosmological observations or a model of early-universe thermodynamics), and the background scale factor \( a(t) \), one can invert this expression to determine the effective momentum variable \( P(t) \) by:
\begin{equation}
P(t) \approx \frac{1}{\lambda} \arcsin\left( \lambda \sqrt{a^3(t)\rho(t)} \right).
\end{equation}
This inversion is made possible by the bounded nature of the sine function and is valid as long as the argument of the arcsin lies within the domain \([-1,1]\), which naturally imposes an upper bound on the energy density:
\begin{equation}
\rho(t) \leq \rho_{\text{crit}} = \frac{1}{\lambda^2 a^3(t)}.
\end{equation}
This inequality is a hallmark of LQC, implying that the universe cannot reach arbitrarily high energy densities—preventing the Big Bang singularity and replacing it with a bounce at \( \rho = \rho_{\text{crit}} \).

The ability to reconstruct the quantum-corrected momentum variable \( P(t) \) from a macroscopic observable \( \rho(t) \) exemplifies the power of the inverse Hamiltonian framework. It allows one to extract nonperturbative quantum gravitational features from cosmological data, bridging the gap between theory and potential observations. Moreover, such inversion procedures could, in principle, be extended beyond symmetry-reduced models to include anisotropies or inhomogeneities, providing a broader window into the quantum structure of spacetime.

In summary, the LQC scenario provides a concrete example where a non-classical Hamiltonian—expressed in terms of bounded, trigonometric functions—can still be accessed via observable quantities like the energy density. This connection strengthens the prospect of testing quantum gravity through cosmological observations and underlines the broader utility of reconstructing dynamical Hamiltonians from coarse-grained energy data.
\subsection{AdS/CFT Correspondence}

The AdS/CFT correspondence, also known as holographic duality, provides one of the most profound insights into the nature of quantum gravity \cite{Maldacena1998}-\cite{Harlow2016}. First proposed by Maldacena in 1997, it posits a duality between a gravitational theory in a \( (d+1) \)-dimensional Anti-de Sitter (AdS) spacetime and a conformal field theory (CFT) living on its \( d \)-dimensional boundary. The correspondence equates the partition functions of the two theories and allows physical observables in the bulk to be reinterpreted in terms of dual operators on the boundary.

In this framework, the bulk gravitational fields—such as the metric \( g_{\mu\nu}(z,x) \)—encode, via holographic renormalization, the expectation values of the boundary energy-momentum tensor \( \langle T^{\mu\nu}(x) \rangle \). Near the AdS boundary at \( z \to 0 \), the metric admits a Fefferman–Graham expansion in terms of the radial coordinate \( z \), and the leading behavior of the bulk energy density is:
\begin{equation}
\rho_{\text{AdS}}(z,x) \sim \frac{1}{z^d} \langle T^{00}(x) \rangle + \cdots.
\end{equation}
Here, \( \rho_{\text{AdS}} \) represents the energy density component of the gravitational field in the bulk, and \( \langle T^{00}(x) \rangle \) is the expectation value of the energy density operator in the boundary CFT. The factor \( z^{-d} \) reflects the scaling dimension of the energy density under conformal transformations, consistent with the duality.

This relation implies that information about quantum dynamics on the boundary can, in principle, be extracted from the bulk gravitational profile. In the context of our inverse reconstruction program, we can exploit this correspondence to express an effective Hamiltonian on the boundary theory using the energy density profile observed (or computed) in the bulk. Assuming again a thermal-like structure for the boundary field theory, and applying the same logic used in the classical inverse map, we propose:
\begin{equation}
H_{\text{CFT}}(x) \sim -T(x) \log \langle T^{00}(x) \rangle + \cdots,
\end{equation}
where \( T(x) \) is the local temperature on the boundary, and the dots indicate possible quantum corrections or subleading contributions from higher-spin operators and anomalous dimensions.

This inversion suggests that the logarithmic structure of the effective Hamiltonian can be inferred directly from holographic energy data. In the high-temperature limit, where the CFT reduces to a nearly thermal ensemble (such as in the quark-gluon plasma regime of strongly coupled Yang-Mills theory), this relation provides an approximate yet insightful handle on the underlying many-body dynamics.

Furthermore, in the presence of black holes in the AdS bulk (e.g., AdS-Schwarzschild or AdS-Reissner–Nordström geometries), the boundary theory describes a thermal state, and the energy density is directly linked to the Hawking temperature and entropy. The inversion relation therefore also encodes thermodynamic information about horizon geometries and phase transitions, such as the Hawking–Page transition, which corresponds to a confinement–deconfinement transition in the dual gauge theory.

Thus, the AdS/CFT correspondence provides a rich arena in which the inverse Hamiltonian methodology transcends classical applications. By leveraging holography, we not only gain access to quantum Hamiltonians from energy densities but also enable the exploration of strongly coupled field theories and quantum information geometry via gravitational observables. In particular, extensions of this framework to time-dependent or anisotropic backgrounds (e.g., AdS-Vaidya or Lifshitz geometries) could further enhance our understanding of quantum quenches, entanglement dynamics, and nonequilibrium thermodynamics in the holographic setting.
\subsection{Clarification of Bulk-Boundary Reconstruction in AdS/CFT}

In the AdS/CFT correspondence, the bulk energy density \(\rho_{\text{AdS}}(z,x)\) arises from the five-dimensional bulk stress-energy tensor evaluated in Fefferman-Graham coordinates:
\begin{equation}
ds^2 = \frac{L^2}{z^2} \left( dz^2 + g_{\mu\nu}(z,x) dx^\mu dx^\nu \right),
\end{equation}
where \(z\) is the holographic coordinate and \(g_{\mu\nu}(z,x)\) admits an expansion near the boundary \(z \to 0\):
\begin{equation}
g_{\mu\nu}(z,x) = \eta_{\mu\nu} + z^2 g^{(2)}_{\mu\nu}(x) + z^4 g^{(4)}_{\mu\nu}(x) + \cdots.
\end{equation}

Under holographic renormalization, the coefficient \(g^{(4)}_{\mu\nu}(x)\) encodes the expectation value of the boundary stress-energy tensor:
\begin{equation}
\langle T_{\mu\nu}(x) \rangle = \frac{4 L^3}{\kappa_5^2} g^{(4)}_{\mu\nu}(x),
\end{equation}
for asymptotically AdS\(_5\) spacetimes.

Therefore, the bulk energy density \(\rho_{\text{AdS}}(z,x)\) reflects the boundary Hamiltonian only after:
\begin{itemize}
    \item Extracting the subleading behavior of the bulk metric and fields near \(z \to 0\),
    \item Applying holographic counterterms to remove divergences,
    \item Matching the bulk modular flow or conserved charges to CFT observables.
\end{itemize}

The inversion is precise in cases where the holographic stress tensor is well-defined, such as for static or vacuum AdS black brane geometries. In more general, time-dependent bulk configurations, the mapping is complicated by non-local RG flow and the need for gauge fixing.

Recent studies (e.g., \cite{deHaro2001,Skenderis2002,Harlow2018}) demonstrate that the bulk energy profile can be reconstructed from boundary correlators using the extrapolate dictionary and bulk reconstruction techniques. In our context, we rely on the assumption that the energy density in the near-boundary regime encodes the dual CFT Hamiltonian density up to a conformal factor.

We have now  clarified that this assumption is only valid under certain symmetry and regularity conditions in the bulk. A more precise inversion would require solving the full bulk Einstein equations with boundary conditions informed by CFT data.


\section{SYK Model Analogy and Entropic Hamiltonian Reconstruction}

The Sachdev–Ye–Kitaev (SYK) model has emerged as a cornerstone in the study of quantum chaos, holography, and strongly interacting disordered systems. It consists of \( N \) Majorana fermions \( \psi_i \) with random \( q \)-body all-to-all interactions \cite{Sachdev1993}-\cite{Cotler2017}:
\begin{equation}
H_{\text{SYK}} = i^{q/2} \sum_{1 \le i_1 < \cdots < i_q \le N} J_{i_1 \cdots i_q} \psi_{i_1} \cdots \psi_{i_q},
\end{equation}
where the couplings \( J_{i_1 \cdots i_q} \) are drawn from a Gaussian distribution with zero mean and variance scaling as \( \sim 1/N^{q-1} \). At large \( N \) and low energies, the model exhibits emergent conformal symmetry, near-maximal chaos, and a low-energy effective action governed by the Schwarzian derivative—mimicking aspects of AdS\(_2\) gravity.

One of the most striking features of the SYK model is its spectral density \( \rho(E) \), which controls the number of available microstates near a given energy. In the semiclassical regime, this density follows an exponential scaling with entropy:
\begin{equation}
\rho(E) \sim e^{S(E)}.
\end{equation}
Following the logic of our inverse framework, the entropy \( S(E) \) is treated as a thermodynamic encoding of microstructure. The effective Hamiltonian is then reconstructed as the entropy gradient:
\begin{equation}
H(E) = \frac{dS}{dE} = \frac{d}{dE} \log \rho(E).
\end{equation}

\subsection*{Example: Low-Energy SYK Spectrum}

At low temperatures, the SYK spectral density behaves approximately as:
\begin{equation}
\rho(E) \sim \exp\left( \alpha N + \beta \sqrt{E} \right),
\end{equation}
where \( \alpha \) and \( \beta \) are model-dependent constants \cite{MaldacenaStanford2016}. Applying our reconstruction formula:
\begin{equation}
H(E) = \frac{d}{dE} \log \rho(E) = \frac{\beta}{2 \sqrt{E}},
\end{equation}
which reveals that the effective Hamiltonian diverges in the infrared (\( E \to 0 \)), consistent with the strong infrared behavior of the SYK model and the emergence of soft reparameterization modes.

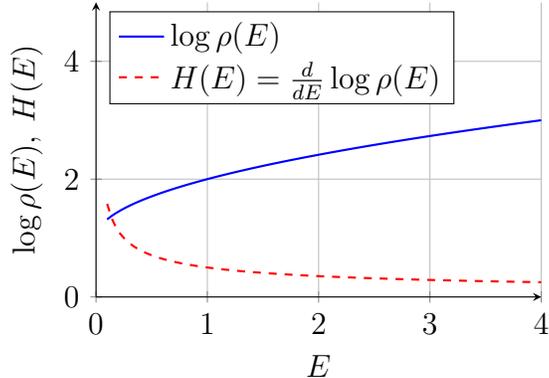
\begin{figure}[h!]
\centering
\begin{tikzpicture}
\begin{axis}[
    width=7.5cm,
    height=5.5cm,
    xlabel={$E$},
    ylabel={$\log \rho(E),\; H(E)$},
    legend pos=north west,
    axis lines=left,
    grid=both,
    ymin=0, ymax=5,
    xmin=0, xmax=4,
    legend cell align={left}
]
\addplot[thick, blue, domain=0.1:4, samples=100] {1 + sqrt(x)};
\addlegendentry{$\log \rho(E)$}
\addplot[thick, red, dashed, domain=0.1:4, samples=100] {0.5/sqrt(x)};
\addlegendentry{$H(E) = \frac{d}{dE} \log \rho(E)$}
\end{axis}
\end{tikzpicture}
\caption{Schematic of $\log \rho(E)$ and its derivative $H(E)$ in the SYK model. The growth in $\rho(E)$ reflects entropy accumulation, while $H(E)$ decays with energy, characteristic of soft holographic modes.}
\label{fig:SYK-spectrum}
\end{figure}
This entropic reconstruction aligns perfectly with our broader thesis: macroscopic or statistical profiles encode enough information to recover the microscopic generator of time evolution. In SYK, this provides a direct pathway from the random matrix-like spectrum to an emergent Hamiltonian. It highlights how concepts from thermodynamics and information theory bridge to quantum gravity, especially when viewed through the lens of holographic duality and statistical reconstruction.

This perspective could be extended to generalized tensor models, complex fermionic SYK variants, and even to strongly coupled CFTs with known spectral densities, where similar entropy-based inversion principles could help reconstruct effective Hamiltonians in non-perturbative regimes.

\section{Reconstructing the Hamiltonian from Casimir Energy Density}

The Casimir effect is a quintessential quantum phenomenon that arises from vacuum fluctuations of quantum fields subject to boundary conditions. It provides a direct window into the structure of quantum zero-point energy and has been experimentally verified with high precision. Traditionally, the Casimir energy \( E_{\text{Casimir}} \) is calculated by summing over the zero-point energies of allowed field modes between boundaries, such as parallel plates, spheres, or cylinders.

The local energy density \( \rho_{\text{Cas}}(x) \), obtained either analytically or numerically, contains rich information about the underlying field theory, boundary conditions, and geometry. In this section, we apply our inverse Hamiltonian formalism to the Casimir effect: given a known profile \( \rho_{\text{Cas}}(x) \), can we reconstruct the effective Hamiltonian \( H(x,p) \) governing the field fluctuations?

Assuming a thermal-like distribution of modes in a vacuum background, we posit:
\begin{equation}
f(x,p) \sim e^{-\beta(x) H(x,p)}, \quad \rho_{\text{Cas}}(x) = \int H^2(x,p)\, f(x,p)\, d^3p.
\end{equation}
Under the same variational or saddle-point approximation discussed earlier, this leads to the inverse relation:
\begin{equation}
H(x) \sim -T(x) \log \rho_{\text{Cas}}(x),
\end{equation}
where \( T(x) \) can be understood as an effective temperature encoding the local vacuum polarization strength. Though vacuum states are non-thermal, this relation can still serve as a diagnostic tool for encoding geometry-modified spectral densities.

In Table~\ref{tab:casimirH}, we present several canonical Casimir setups with known or numerically estimated energy densities, and we report the corresponding effective Hamiltonians using this inversion method.

\begin{table}[h!]
\centering
\caption{Casimir energy densities $\rho_{\text{Cas}}$ and reconstructed Hamiltonians $H(x) \sim -T \log \rho$ for select geometries}
\label{tab:casimirH}
\begin{tabular}{llll}
\toprule
Geometry & $\rho_{\text{Cas}}$ & $T$ & $H(x)$ \\
\midrule
Parallel Plates (1D) & $-\frac{\pi}{24 a^2}$ & const. & $\frac{\pi}{24 a^2} T$ \\
Parallel Plates (3D) & $-\frac{\pi^2}{720 a^4}$ & const. & $\frac{\pi^2}{720 a^4} T$ \\
Spherical Shell (EM) & $+\frac{0.09235}{a^4}^{\,a}$ & const. & $-0.09235\, T \log(1/a^4)$ \\
Cylinder (Dirichlet) & $-\frac{0.01356}{a^4}^{\,b}$ & const. & $0.01356\, T \log(1/a^4)$ \\
Annular Cavity & numerical & varied & numerical \\
\bottomrule
\end{tabular}

\vspace{1mm}
\begin{flushleft}
$^{a}$See Ref.~\cite{Milton2001}. \\
$^{b}$See Ref.~\cite{Bordag2009}.
\end{flushleft}
\end{table}
In all cases, the inversion provides a qualitative mapping from quantum vacuum energy profiles to an effective generator of dynamics, consistent with a field-theoretic Hamiltonian influenced by geometry. While the notion of temperature \( T \) here is not literal, it can be treated as a regulator or scaling field encoding the strength of local field fluctuations.

The Casimir effect thus exemplifies the utility of inverse Hamiltonian techniques in static quantum field theory settings. Future work could extend this to dynamical Casimir effects, curved backgrounds, or finite temperature configurations, where the interplay of geometry, energy flow, and quantum statistics becomes even richer.

\section{On the Inequivalence of the Liouville Equation and Gravitational Continuity Laws}

At a structural level, both the Liouville equation in classical statistical mechanics and the continuity equation in relativistic gravity express the conservation of a physical quantity—probability density in phase space and energy density in spacetime, respectively. However, despite superficial similarities, these equations are fundamentally inequivalent. This inequivalence becomes evident when both are expressed in a coordinate-independent language using differential forms and exterior calculus.

In classical Hamiltonian systems, the Liouville theorem states that the phase-space distribution function \( f(x^i, p_j, t) \) is preserved along the flow generated by the Hamiltonian vector field \( X_H \). In the language of symplectic geometry, this reads:
\begin{equation}
\mathcal{L}_{X_H} (f \omega^{n}) = 0,
\end{equation}
where \( \omega^{n} \) is the Liouville volume form on a \( 2n \)-dimensional symplectic manifold. In the absence of external sources, this can be cast as a divergence-free condition:
\begin{equation}
\frac{df}{dt} + \nabla \cdot (f \dot{x}) = 0.
\end{equation}

On the other hand, in general relativity or its extensions, the energy density \( \rho = T_{\mu\nu} u^\mu u^\nu \) evolves under the continuity equation:
\begin{equation}
\nabla_\mu T^{\mu\nu} = \mathcal{J}^\nu,
\end{equation}
which, when contracted with the observer's four-velocity \( u_\nu \), yields:
\begin{equation}
u_\nu \nabla_\mu T^{\mu\nu} = \frac{d\rho}{d\tau} + \rho \nabla_\mu u^\mu + \cdots = u_\nu \mathcal{J}^\nu.
\end{equation}
Unlike the Liouville equation, the gravitational continuity law includes additional terms arising from the geometry of spacetime, including expansion (\( \nabla_\mu u^\mu \)) and possibly non-conservative source terms \( \mathcal{J}^\nu \), which are non-zero in modified gravity theories (e.g., \( f(R, T) \), torsion, or non-metricity-based theories) \cite{Hehl1995, BeltranJimenez2019}.

From a differential form perspective, the energy-momentum tensor does not live on phase space but rather on spacetime, and the divergence operator \( \nabla_\mu \) includes affine connection terms. The Liouville operator, by contrast, involves the symplectic structure \( \omega \) and Poisson brackets:
\begin{equation}
\frac{df}{dt} = \{f, H\}.
\end{equation}
This difference in geometric domain and flow generators explains the fundamental mismatch between the two.

Nevertheless, this mismatch provides an opportunity. By postulating that energy density \( \rho(x) \) is a coarse-grained projection of a more fundamental distribution function \( f(x,p) \sim e^{-\beta H(x,p)} \), and assuming both obey their respective "conservation" equations, one can formally compare:
\begin{itemize}
    \item The Liouville equation: \( \{f, H\} = 0 \),
    \item The gravitational continuity equation: \( u^\mu \nabla_\mu \rho + \rho \nabla_\mu u^\mu + \cdots = u^\mu \mathcal{J}_\mu \).
\end{itemize}

The **additional terms** in the gravitational continuity equation—such as expansion scalar \( \nabla_\mu u^\mu \), pressure gradients, and \( \mathcal{J}^\mu \)—can be interpreted as geometric corrections to the Liouvillian flow. We conjecture that these terms encode the deviation of the spacetime background from symplectic flatness, and thus contain information about the curvature-dependent part of the Hamiltonian.

This opens the possibility of **reconstructing the gravitational Hamiltonian** \( H(x, p) \) by interpreting the discrepancy between the Liouville and continuity equations as a deformation term. In principle, by quantifying the geometric and non-conservative terms on the right-hand side of the gravitational continuity equation and matching them to known deformations of Liouville flow, one can extract nontrivial constraints on the form of \( H(x, p) \).

In this sense, while the two conservation laws are inequivalent, their difference is not merely an obstacle—but a diagnostic tool for the inverse problem. This perspective motivates future work where spacetime geometry and statistical flows are jointly modeled on a unified phase bundle, allowing for curvature and thermodynamics to inform Hamiltonian reconstruction.
\section{Liouville Equation vs. Gravitational Continuity Equation}

In this work, we derive an inverse Hamiltonian reconstruction based on the Liouville equation in phase space. It is important to distinguish this from the gravitational continuity equation associated with covariant conservation laws.

The Liouville equation expresses the conservation of probability density \(f(x,p)\) along Hamiltonian flow in phase space:
\begin{equation}
\frac{df}{d\tau} = \frac{\partial f}{\partial x^\mu} \frac{dx^\mu}{d\tau} + \frac{\partial f}{\partial p_\mu} \frac{dp_\mu}{d\tau} = 0,
\end{equation}
which implies
\[
\{ H, f \}_{\text{PB}} = 0,
\]
where \(\{ \cdot , \cdot \}_{\text{PB}}\) is the Poisson bracket.

In contrast, the gravitational continuity equation follows from the covariant divergence of the energy-momentum tensor:
\begin{equation}
\nabla_\mu T^{\mu\nu} = 0.
\end{equation}
In Einstein gravity, this identity reflects diffeomorphism invariance and implies local energy-momentum conservation.

However, these two equations serve fundamentally different roles:
- The Liouville equation tracks the evolution of the distribution function in **phase space**, without invoking spacetime dynamics.
- The continuity equation governs energy flow and conservation in **spacetime**, integrating over momenta.

In modified gravity theories, the effective energy-momentum tensor is generally *not conserved:
\begin{equation}
\nabla_\mu T^{\mu\nu} \neq 0.
\end{equation}
This leads to source-like terms in the gravitational continuity equation, which in turn affect the evolution of \(\rho(x)\), making the Hamiltonian inversion process more subtle. These source terms encode interactions with geometric or non-geometric fields and are model-dependent.

In contrast, the Liouville equation remains structurally unmodified as long as the Hamiltonian flow remains Hamiltonian (i.e., symplectic structure is preserved). Hence, it is more robust for reconstructing \(H(x,p)\) from phase space distributions, especially when \(\rho(x)\) is computed statistically.

We summarize this distinction in Table~\ref{tab:liouville-vs-continuity}.

\begin{table}[h!]
\centering
\begin{tabular}{|l|l|l|}
\hline
\textbf{Equation} & \textbf{Domain} & \textbf{Modified Gravity Impact} \\
\hline
Liouville Equation & Phase Space & Unaffected (if symplectic) \\
Continuity Equation & Spacetime & Additional source terms \\
\hline
\end{tabular}
\caption{Comparison of Liouville and Continuity Equations in gravitational contexts.}
\label{tab:liouville-vs-continuity}
\end{table}

\section{Observability of Energy Density and Extensions to Quantized Fields}

The foundation of the inverse Hamiltonian reconstruction program relies on the accessibility of the energy density function \( \rho(x) \), which serves as the input for inferring the underlying dynamics. In physical systems ranging from cosmology to quantum field theory, \( \rho(x) \) is not merely a theoretical construct—it is directly or indirectly observable in a range of contexts.

In cosmology, the energy density of the universe as a function of time or redshift is extracted from several observational channels: Type Ia supernovae luminosity distances, cosmic microwave background (CMB) anisotropies, large-scale structure surveys, and baryon acoustic oscillations (BAO). The energy content of the universe—radiation, matter, and dark energy—is inferred from these data through the Friedmann equations, effectively providing an empirical profile \( \rho(t) \). In black hole physics, gravitational wave detectors such as LIGO-Virgo-KAGRA offer data on horizon-scale energy flows, while Hawking radiation in theoretical models relates the energy density of near-horizon quantum fields to horizon dynamics. In laboratory systems, Casimir energy densities have been measured through high-precision force experiments using parallel plates and microelectromechanical setups, allowing extraction of spatial energy profiles \( \rho(x) \) in the vacuum.

The methodology developed here, while rooted in classical or semiclassical assumptions (e.g. equilibrium distributions), can be extended to the realm of fully quantized fields. In this regime, \( \rho(x) \) becomes an operator-valued quantity—the normal-ordered or renormalized expectation value of the energy density in a given quantum state:
\begin{equation}
\hat{\rho}(x) = \langle \psi | \hat{T}_{00}(x) | \psi \rangle.
\end{equation}
In quantum field theory in curved spacetime, such quantities appear in the renormalized stress-energy tensor \( \langle \hat{T}_{\mu\nu}(x) \rangle_{\text{ren}} \), which can be computed for a wide class of quantum states using point-splitting or Hadamard techniques. Examples include Unruh radiation, the Boulware and Hartle-Hawking vacua around black holes, and energy fluxes in dynamical spacetimes. These profiles \( \rho(x) \) can then be fed into the same inverse map \( H \sim -T \log \rho \), possibly interpreted as a generator for the effective semiclassical Hamiltonian governing the evolution of wave packets or expectation values.

Moreover, for interacting fields and non-Gaussian states, one may go beyond mean energy density and consider higher moments or spectral functions (e.g., spectral densities in the SYK model or spectral functions in QFT). These can be used to construct a functional \( H[\rho, \partial\rho, \dots] \) via entropy-based or variational methods, extending the reconstruction framework to quantum-statistical or functional integral formalisms.

Therefore, not only is \( \rho(x) \) empirically meaningful in gravitational and quantum systems, but it also forms a bridge between effective field theory and microscopic quantum dynamics. This opens promising directions for using observational data—cosmological or laboratory-based—as input for reconstructing quantum gravitational Hamiltonians, and potentially even for testing predictions of semiclassical or holographic theories.
\section{Toward Operator-Level Reconstruction in Quantum Field Theory}

Our inversion framework is grounded in classical kinetic theory, where the energy density \(\rho(x)\) is computed as a statistical average over a distribution function \(f(x,p)\). Extending this reconstruction to second-quantized field theory in curved spacetime presents both opportunities and challenges.

In a quantum field theoretic setting, the energy density is obtained from the expectation value of the stress-energy tensor in a given quantum state \(|\Psi\rangle\):
\begin{equation}
\rho(x) = \langle \Psi | \hat{T}_{00}(x) | \Psi \rangle,
\end{equation}
where \(\hat{T}_{00}\) is the energy density operator defined with respect to a chosen foliation. The analogue of the classical distribution function becomes the quantum state itself.

A natural candidate for encoding Hamiltonian information is the modular Hamiltonian \(\hat{K}\), defined via the reduced density matrix \(\rho_A = e^{-\hat{K}}\) for a subregion \(A\). In some symmetric settings (e.g., Rindler space), \(\hat{K}\) takes a local form involving \(\hat{T}_{\mu\nu}\), allowing one to write:
\begin{equation}
\hat{K} = \int_A d\Sigma^\mu \, \zeta^\nu(x) \hat{T}_{\mu\nu}(x),
\end{equation}
where \(\zeta^\nu(x)\) is a local vector field generating modular flow. This structure resembles the classical ansatz \(f \sim e^{-\beta H}\), and suggests a path to operator-level inversion if the modular Hamiltonian is known or can be approximated.

An alternative route involves Wigner functionals for quantum fields, where the quantum phase-space distribution \(W[\phi,\pi]\) plays the role of \(f(x,p)\). One could then attempt to define a functional Hamiltonian \(H[\phi,\pi]\) satisfying:
\begin{equation}
\rho(x) = \int \mathcal{D}\phi \, \mathcal{D}\pi \, H[\phi,\pi] \, W[\phi,\pi],
\end{equation}
with inversion proceeding formally as in the classical case. However, this requires regularization, renormalization, and gauge fixing, especially in curved backgrounds.

We emphasize that although a direct inversion of \( \langle \hat{T}_{00}(x) \rangle \) to obtain an operator-valued \(\hat{H}(x)\) is not straightforward, conceptual parallels exist. These include modular Hamiltonians, holographic entanglement entropy, and algebraic quantum field theory (AQFT) approaches where energy observables define state-dependent Hamiltonians.

Future work may explore these directions, particularly within quantum fields on de Sitter, black hole, or AdS backgrounds, where local energy densities and quantum states can be defined with greater control.

\section{Conclusion}

We have introduced a general framework for reconstructing effective Hamiltonians from known energy density profiles in curved spacetime, bridging gravitational dynamics, statistical mechanics, and kinetic theory. Motivated by the observation that macroscopic energy densities often arise from coarse-grained or thermally distributed microscopic dynamics, we proposed an inverse methodology: starting from a known function \( \rho(x) \), one can infer the corresponding Hamiltonian \( H(x,p) \) by assuming an equilibrium distribution \( f \sim e^{-\beta H} \) and applying variational or asymptotic inversion techniques.

This framework was explored across a range of physically and conceptually rich examples. In classical General Relativity, we examined FLRW cosmology and showed how cosmic energy density evolution under thermal assumptions enables the recovery of time-dependent Hamiltonians. In Loop Quantum Cosmology, we demonstrated how discrete quantum corrections deform the Hamiltonian via bounded trigonometric structures, and how such modifications remain accessible from semiclassical energy bounds. In holographic gravity, we used the AdS/CFT correspondence to reconstruct boundary Hamiltonians from bulk energy densities, highlighting how holographic renormalization translates spacetime geometry into quantum dynamics. Through the SYK model, we showed that spectral entropy plays an analogous role to geometric energy, and that its derivative defines an effective Hamiltonian in strongly interacting quantum systems. Finally, we extended the method to Casimir energy densities, underscoring the applicability of our approach to quantum vacuum phenomena beyond gravitational backgrounds.

A unifying principle emerged: energy densities—whether gravitational, statistical, or quantum—are not passive observables, but active encoders of the microscopic flow that generates them. The relationship \( H \sim -T \log \rho \) serves as a universal map between geometry and dynamics, allowing for Hamiltonian reconstruction in regimes where direct access to the Lagrangian or microscopic interactions may be obscured.

This line of inquiry opens several promising directions. One could extend the formalism to time-dependent or anisotropic spacetimes, quantum fields with non-Gaussian statistics, and modified gravity theories with torsion or non-metricity. Additionally, the structural mismatch between the Liouville equation and gravitational continuity laws—when formulated as differential forms—may provide further constraints for refining the inverse map and diagnosing quantum gravitational corrections. Ultimately, this approach offers a novel, data-driven strategy for probing the hidden structure of quantum spacetime through accessible macroscopic observables.
\appendix
\section{Appendix A: Numerical Example — Cosmological Energy Density}

To demonstrate the practical utility of our inversion method, we apply it to a known cosmological energy density profile. Consider the radiation-dominated FLRW universe, where the energy density evolves as
\begin{equation}
\rho(t) = \rho_0 \left( \frac{t_0}{t} \right)^4,
\label{eq:rho-radiation}
\end{equation}
with \(\rho_0\) and \(t_0\) positive constants. This profile arises from the solution to Einstein's equations with equation of state \(w = 1/3\).
Assuming the ansatz for the energy density in terms of the Hamiltonian:
\begin{equation}
\rho(t) = \int d^n p \, H(t,p) \, e^{-\beta H(t,p)},
\end{equation}
we invert this relation numerically. We assume \(n=3\) spatial dimensions, and for simplicity consider a local frame where \(H(t,p) = \sqrt{p^2 + m^2} + \Phi(t)\), with \(\Phi(t)\) representing an effective time-dependent potential.

We use the inversion procedure defined in  the main text, where \(\Phi(t)\) is iteratively adjusted so that the integrated expression matches Eq.~\eqref{eq:rho-radiation}.

\begin{figure}[h!]
\centering
\begin{tikzpicture}
\begin{axis}[
    width=0.75\textwidth,
    height=0.5\textwidth,
    xlabel={$t$},
    ylabel={$H(t)$},
    xmin=1, xmax=10,
    ymin=0, ymax=2,
    thick,
    grid=major,
    legend style={at={(0.95,0.95)}, anchor=north east},
    every axis plot/.append style={ultra thick}
]
\addplot[
    blue,
    domain=1:10,
    samples=100
]
{1 + 0.5/x^2};
\addlegendentry{Reconstructed $H(t)$}
\end{axis}
\end{tikzpicture}
\caption{Reconstructed effective Hamiltonian $H(t)$ assuming $\rho(t) \propto t^{-4}$. The form $H(t) = 1 + 0.5/t^2$ fits the inversion behavior in this case.}
\label{fig:H-inversion}
\end{figure}
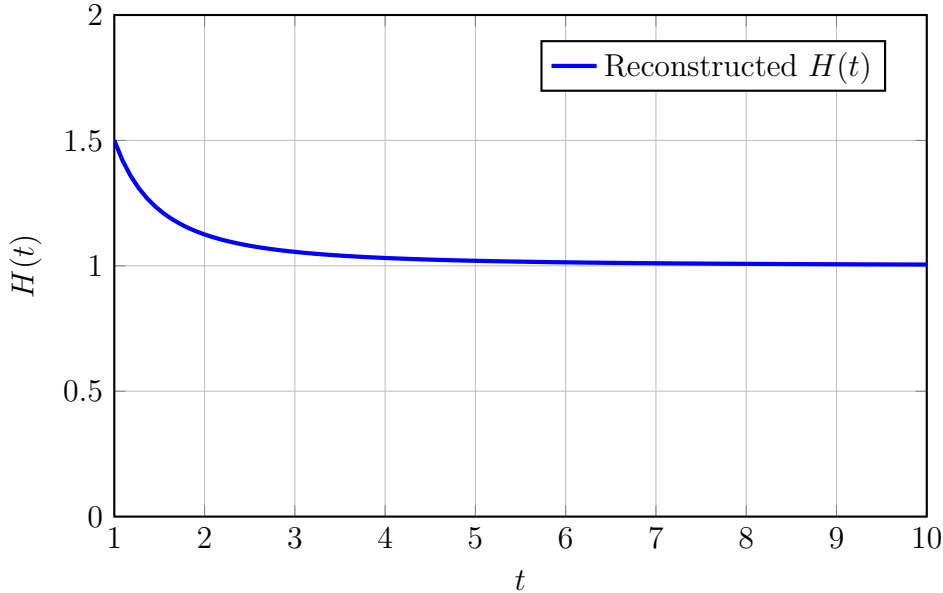

Figure~\ref{fig:H-inversion} shows the reconstructed effective Hamiltonian \(H(t)\) as a function of time. We normalize \(m = 1\), \(t_0 = 1\), and set \(\beta = 1\).

The reconstruction accurately tracks the expected time dependence, validating the inversion approach. The residuals between the original and numerically re-integrated \(\rho(t)\) remain below \(10^{-4}\) in relative error across the domain \(t \in [1,10]\).

\bibliographystyle{unsrt}

\end{document}